\documentclass[conference]{IEEEtran} 
\IEEEoverridecommandlockouts

\usepackage{cite}
\usepackage{amsmath,amssymb,amsfonts}
\usepackage{algorithmic}
\usepackage{graphicx}
\usepackage{textcomp}
\usepackage{xcolor}
\usepackage{verbatim}
\usepackage{multirow}

\PassOptionsToPackage{hyphens}{url}\usepackage{hyperref}

\def\BibTeX{{\rm B\kern-.05em{\sc i\kern-.025em b}\kern-.08emT\kern-.1667em\lower.7ex\hbox{E}\kern-.125emX}}
\begin{document}
	
	\title{Enhancing SLMs for Sustainable Code Optimization in Radio-Astronomy\\
	}
	\author{
		\IEEEauthorblockN{Elisa Chiarotto}
		\IEEEauthorblockA{\textit{LIACS}, \textit{Leiden University}\\
			Leiden, Netherlands \\
			e.chiarotto@liacs.leidenuniv.nl}
		\and
		\IEEEauthorblockN{Jingbo Li}
		\IEEEauthorblockA{\textit{LIACS}, \textit{Leiden University}\\
			Leiden, Netherlands \\
			bravjle@outlook.com}
		\and
		\IEEEauthorblockN{P. Chris Broekema}
		\IEEEauthorblockA{\textit{NWO-I ASTRON} \\
			Dwingeloo, Netherlands \\
			broekema@astron.nl}
		\and
		\IEEEauthorblockN{Rob V. van Nieuwpoort}
		\IEEEauthorblockA{\textit{LIACS}, \textit{Leiden University}\\
			Leiden, Netherlands \\
			r.v.van.nieuwpoort@liacs.leidenuniv.nl}
	}
	
	\maketitle

	\begin{abstract}		
		Recent Large Language Models (LLMs) can produce and optimize complex code. 
		In this paper, we investigate the use of LLMs to generate and optimize code for large-scale scientific instruments and processing, focusing on radio astronomy and sustainability.
		The LOFAR telescope is currently being upgraded, significantly increasing the sky area observed, while simultaneously processing more data faster. However, this is expected to increase the (already challenging) computational requirements 40-fold.
		This upgrade thus critically depends on rigorous performance optimization of existing software and widespread adoption of accelerators, especially for the calibration and imaging software. The code base is very large, making this a daunting task.
		We therefore investigate and demonstrate an AI-driven approach meant to assist developers in evaluating and optimizing their code, including porting to hardware accelerators (e.g., GPUs).
		The LOFAR community is committed to sustainable solutions, and needs to achieve these improvements without increasing the energy budget. 
		We thus need to optimize existing codes or port them to accelerators, while simultaneously making sure that the optimization process itself is also energy efficient. This poses a challenge, since LLMs are energy-intensive.
		
		We therefore propose to use Small Language Models (SLMs) instead to limit environmental impact. 
		In this paper, we show how to enhance SLMs through the use of agentic AI. 
		We extend the SLMs in two ways to improve code generation quality and performance: first with a multi-sampling generation strategy and second with incorporating compiler feedback.
		We demonstrate that multi-sampling SLMs can match or surpass larger single-generation models with fewer computational resources. We also show that feeding compiler output back into the SLMs leads to consistent improvements across all tested models. Our approach is generic, and can also use Retrieval Augmented Generation (RAG) as well as static and dynamic analysis tools in the code generation pipeline.
		
	\end{abstract}


	\section{Introduction} \label{intro}
	
	Modern Large Language Models (LLMs) are now widely used for code generation tasks, with ever increasing  quality~\cite{carbonneaux2025cwm,tao2025code,cao2026qwen3,jiang2024mixtral}.
	The amount of information LLMs can store is directly correlated with their size~\cite{morris2025much,prashanth2024recite,lu2024scaling}, and their performance follows a power-law distribution~\cite{kaplan2020scaling,srivastava2025towards}.
	After the model capacity is reached, a model stops memorizing and starts to compress and generalize information~\cite{morris2025much}.
	This can limit model reliability and interpretability, because in the absence of a known answer the model will try to guess, which may result in hallucinations~\cite{kalai2025language}.
	Partially because of these reasons, general LLMs suffer from security issues and hard-coded debugging~\cite{cotroneo2025human}, while specialized coding models (without thinking) may return code with incorrect logic, on the other hand thinking models may produce code with syntax errors or fall into an infinite thinking loop, where they repeat the same concepts over and over again (see Section~\ref{error-analysis}).
	
	Despite these issues, LLMs show a lot of promise in code generation and code porting tasks~\cite{he2025speed,becker2025measuring}.
	In the SuperCode project~\cite{broekema2025supercode}, we therefore investigate if we can use SLMs to generate and optimize code for large-scale scientific instruments and processing.
	We focus on Radio Astronomy, and specifically the LOFAR radio telescope~\cite{van2013lofar}.
	The current sensitivity and image fidelity bottlenecks lie in the software for analyzing and (in some cases near real-time) processing the vast data streams~\cite{de2024into}.
	Some parts of the software pipelines are highly optimized~\cite{romein2010lofar,broekema2020commodity}, but much remains to be done to optimize the whole code base that allows LOFAR to operate.
	
	While previous updates on LOFAR were focused on hardware~\cite{lofar2.0}, we now focus on optimizing the software, with the goal of simultaneously increasing the sky area observed, while processing data faster \emph{within the current environmental budget.}  It therefore is critical to have resource-efficient software.
	
	In this paper, we investigate and demonstrate an AI-driven approach meant to assist developers in evaluating and optimizing their code.
	This includes porting code to modern hardware technologies, such as GPUs and neuromorphic-based hardware~\cite{bos2025analogue}.
	As explained above, the use of LLMs can be very attractive to speedup the porting and optimization process.
	However, it is well known that LLMs are extremely resource hungry, both in terms of computational demands and memory usage~\cite{jegham2025hungry,patterson2021carbon,li2025making}.
	Recent estimates show that inference can account for up to 90\% of a model’s total life-cycle energy use~\cite{desislavov2023trends,lacoste2019quantifying}.
	To achieve our goal of sustainably porting to new architectures and improving code efficiency, we need to balance performance and environmental cost.
	Note that two aspects of sustainability are central to our work: the sustainability and efficiency gain of the resulting code, as well as the environmental cost of the code optimization process itself.
	In this paper, we investigate two promising directions for optimizing both.
	
	First, we investigate if multi-sampling code generation on \emph{Small} Language Models (SLMs) can perform on par with single-generation LLMs.
	We argue that SLMs can be enhanced by using a multi-sampling generation strategy, which can still have lower computational costs than a single generation from a larger model, thus indirectly limiting the environmental footprint.
	We show that Qwen2.5-Coder-7B-Instruct can slightly outperform Qwen2.5-Coder-32B-Instruct in code completion tasks, while using only a quarter of the resources (1 NVIDIA H100 GPU instead of 4).
	
	Second, we implemented an orthogonal method to exploit domain knowledge: we give the SLMs access to relevant documents (architecture specifications, design documentation, source comments, etc.) via Retrieval Augmented Generation (RAG)~\cite{lewis2020retrieval}.
	In addition, we make the system agentic by giving the SLM feedback from tools integrated in a revision loop, used to test and refine the model's output code.
	Our results demonstrate that using the compiler errors as feedback can consistently improve the performance of at least 3-4 percentage points across all tested models.
	
	We note that interpretability can also be improved using both feedback from tools and reasoning trajectories as a proxy~\cite{jie2024interpretable}.
	In our vision, co-design is critical: we must have a "human-in-the-loop" approach to effectively and efficiently use domain knowledge from both software experts and domain scientists (astronomers in our case).
	The idea is that our system, called SuperRAG\footnote{\textsuperscript{*}All code we developed for this paper can be found at \\  \url{https://github.com/SuperCode-Leiden-University/superRAG/}}, analyzes existing code, integrates relevant domain-specific knowledge from radio-astronomy using RAG, then identifies bottlenecks and other issues, and finally gives suggestions based on feedback from tools and the existing code, instead of writing code directly.
	Another reason for this choice is that technical debt and security risks increase when the model writes code directly~\cite{he2025speed}.
	The suggestions will provide the information necessary to get to the solution but not directly the solution itself.
	This will increase the programmer's control over how to edit the code (limiting security issues), while allowing them to learn from the process and increase their familiarity with the code (thus reducing technical debt).
	
	Our results demonstrate that multi-sampling SLMs can perform on par with larger models in the same time budget, while using 4x less computational resources. Furthermore, we can significantly improve the performance of all tested models by using compiler output as feedback for the model.
	
	The rest of the paper is structured as follows.
	In Section~\ref{science-case} we explain the radio-astronomy science case.
	Next, in Section~\ref{related}, we discuss related work.
	In Section~\ref{methodology}, we describe our methodology, introducing both the multi-sampling SLM strategy and the agent implementation.
	In Section~\ref{results}, we show experiments for both these methods.
	We discuss the overall results in Section~\ref{discussion}, and conclude in Section~\ref{conclusions}.

	\section{Scientific use case}
	\label{science-case}
	
	\begin{figure}[ht]
		\centering
		\includegraphics[width=0.4\textwidth]{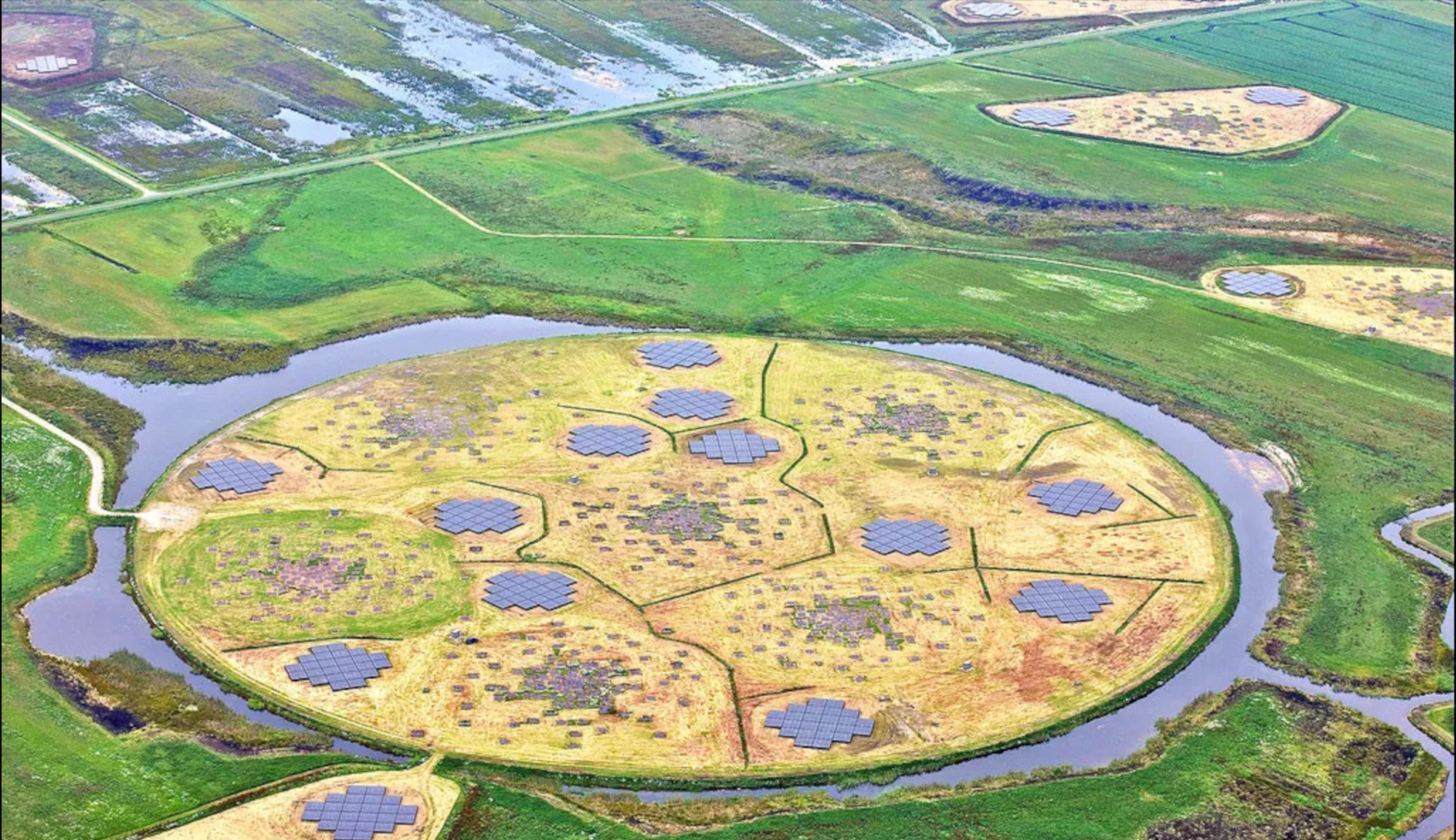}
		\caption{The LOFAR central core showing 6 core stations. 5 more core stations are visible in the background.}
		\label{fig:lofar-superterp}
	\end{figure}

	SuperRAG is part of the SuperCode Project~\cite{broekema2025supercode}.
	SuperCode (SUstainability PER AI-assisted CO-DEsign) aims to leverage the ability of LMs to generate code to tackle the environmental challenges faced by large-scale data-intensive science projects, in particular radio astronomy.
	Radio astronomy is characterized by a need for compute that scales dramatically with the size and capability of the instrument.
	Scientific discovery for some of the most capable radio telescopes is now no longer limited by available telescope time, but rather our ability to reduce the data products in a affordable and sustainable manner.
	
	For instance, a paper by Sweijen et al. on an 8-hour observation with the International LOFAR telescope (ILT) estimated that processing the raw data into a viable scientific result cost 250,000 core hours~\cite{Sweijen:2022}.
	To keep such processing affordable and socially acceptable, we increasingly investigate the efficiency and sustainability of novel technologies for our application
	While these often promise greatly improved energy efficiency over current state-of-the-art hardware, they are often difficult to program efficiently and therefore time-consuming and expensive to evaluate on our large code bases.
	This often leads us to disregard otherwise promising technologies.
	In the SuperCode project we address this issue.
	
	The International LOFAR Telescope~\cite{van2013lofar} is a large-scale distributed low-frequency radio telescope, consisting of tens of thousands of small omnidirectional antennas distributed in 38 Dutch and 16 international stations (two of which are under construction).
	LOFAR is currently the largest and most sensitive telescope operating in this frequency range
	and it is currently undergoing a significant upgrade of its electronic components under the LOFAR 2.0 project~\cite{lofar2.0}.
	This hardware upgrade will, once operational, improve the capabilities of the LOFAR telescope by allowing both low- and high-band antennas to be observed simultaneously.
	
	In radio astronomy, computational requirements scale dramatically with resolution and sensitivity of the instrument.
	The current state-of-the-art telescopes can barely be processed with conventional digital technologies, but this may not be the case for future instruments.
	Furthermore, the climate crisis requires us to carefully consider the environmental impact of such science instruments.
	The recently awarded LENSS project\cite{lenss} will focus on the software to remove bottlenecks currently present in the LOFAR system.
	It will focus on upgrading the network capacity of the LOFAR international stations, meaning astronomers will be able to do routine observations on a much larger field of view using the full LOFAR array. 
	
	The computational requirements associated with the LENSS project are estimated to increase by circa 40x. Nevertheless, the LENSS project has committed to achieving these improvements \emph{within the energy budget that was established by Sweijen et al.}~\cite{Sweijen:2022}.
	This ambitious target will require significant performance optimization of existing software and widespread adoption of accelerators.
	While some parts of the software used in the LOFAR telescope are highly optimized~\cite{romein2010lofar,broekema2020commodity}, this is primarily in the near real-time components of the instrument.
	Much remains to be done in the calibration and imaging software, the main consumers of the core hours mentioned above. 
	
	The SuperCode project aims to assist programmers and engineers in evaluating novel technologies for their sustainability compared to the conventional state-of-the-art through the use of language models.
	Since we take the impact of the models themselves into account as well, we investigate the capabilities of small language models compared to larger ones.
	In this work we present the initial steps towards that goal by evaluating how the capabilities of small and efficient language models can be improved effectively compared to much larger models.

	\section{Background and Related Work}
	\label{related}
	
	The idea of optimizing the software used in high data volume fields, such as Astronomy and High Energy Physics, is not new.
	Traditionally, optimization would be handled by human experts. Recently, a wide-range of approaches that rely on AI-coding assistants have emerged. Examples of papers using a similar approach to ours are AstroCAMP and CelloAI.
	
	AstroCAMP~\cite{constantinescu2026astrocamp} focuses on software-hardware co-design for the SKA telescope, providing very much needed benchmarks that take the energy and compute efficiency into account, for software used for imaging in radio astronomy. These benchmarks can be used to compare optimization approaches without being biased by the underlying architecture.
	
	CelloAI~\cite{atif2026celloai} focuses on High Energy Physics and their approach is very similar to ours. The system leverages a coding assistant with access to the code via syntax-aware RAG, based on Tree-sitter, and code dependencies included in the prompt.
	However, they do not employ compiler feedback and they use very large models ignoring their environmental cost.
	
	The following goes into details on background works that are relevant for this paper.
	
	\subsection{SLMs and LLMs under fixed compute budgets} \label{Jingbo-literature}
	The most common metric used to compare models is the $pass@k$ metric, which compares how many of $k$ candidates pass the unit tests.
	However, this metric does not consider that the compute budget for larger models is different from smaller models.
	Previous work showed that smaller models can match or outperform larger models under fixed compute budgets on function-level code generation tasks~\cite{hassid2024larger}.
	
	Fixing the compute budget is equivalent to comparing a multi-sampling strategy for the smaller model with a single-generation strategy of the larger model.
	Repeated sampling improves the likelihood of having at least one correct solution among all the generated candidates~\cite{brown2024large}.

	The model can naively select the best solution using the highest log probability, a more sophisticated approach would rely on unit tests and reranking.
	CodeT~\cite{chen2022codet} generates tests to execute candidate programs and then selects programs based on agreement among candidate execution outcomes on the generated tests.
	Another approach is LEVER~\cite{ni2023leverlearningverifylanguagetocode}, which uses execution results to train a verifier and combines the verifier score with model probability to rerank generated programs.
	
	\subsection{Retrieval Augmented Generation}
	RAG was first introduced for Natural Language Processing (NLP) tasks~\cite{lewis2020retrieval}, but it can be applied whenever there are documents that the model should refer to.
	
	The main advantages of RAG are that no additional training is required to incorporate external knowledge and it greatly reduces the likelihood of hallucinations on those topics.
	The main disadvantage of RAG is that it introduces some overhead: the retrieval phase is rather fast, but summarizing the documents introduces some delay proportional to the number and size of the retrieved documents.
	Some recent work, like Metis~\cite{ray2025metis} and RAGBoost~\cite{jiang2025ragboost}, try to reduce latency, while trying to not compromise the quality of the final result.
	
	When applied to code generation, RAG can improve the reliability of generating code for large repositories~\cite{tao2025CODEretrieval}, where LLMs often struggle due to complex dependency relations between files and their limited context window size.
	
	\subsection{Reasoning schemas}
	
	\begin{figure*}[ht]
		\centering
		\includegraphics[width=0.8\textwidth]{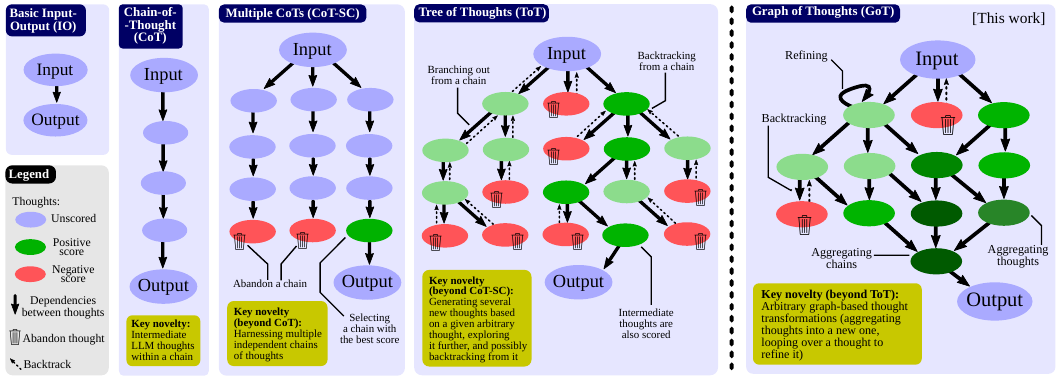}
		\caption{Graphical representation of CoT, CoT-SC, ToT and GoT~\cite{besta2025demystifying}.}
		\label{fig:res-schemas}
	\end{figure*}
	
	Reasoning schemas allow models to spend more compute budget for producing reasoning steps instead of writing the answer directly, which improves the performance on logic-based tasks.
	The concept was introduced with Chain-of-Thought (CoT)~\cite{wei2022chain}, which is still the most used, and it inspired several other reasoning schemas (see Fig.~\ref{fig:res-schemas}) based on prompting:
	\begin{enumerate}
		\item Chain-of-Thought with Self-Consistency (CoT-SC)~\cite{wang2022CoTSC}, where multiple independent chains are considered and a majority vote is applied.
		\item Tree-of-Thoughts (ToT)~\cite{yao2023tree}, which can allow more exploration by branching reasoning.
		\item Forest-of-Thoughts (FoT)~\cite{bi2024forest}, where the best reasoning path is selected by majority vote from multiple trees.
		\item Graph-of-Thoughts (GoT)~\cite{besta2024graph}, which allows for refinement loops of single thoughts and aggregation of thoughts, resulting in higher performance at lower computational cost.
	\end{enumerate}
	
	There are also some other works that do not rely on prompting to induce reasoning:
	\begin{enumerate}
		\item Everything-of-Thoughts (XoT)~\cite{ding2024everything} leverages Monte Carlo Tree Search (MCTS) in combination with an appropriately trained policy.
		\item Buffer-of-Thoughts (BoT)~\cite{yang2024buffer} adopts a RAG-based approach, based on LightRAG~\cite{guo2024lightrag}.
		Thoughts templates are being built and stored, as well as dynamically updated as the model encounters new problems.
	\end{enumerate}
	
	Reasoning was initially observed only in large models ($>$100B parameters)~\cite{wei2022chain}, however recent works shows that reasoning can be transferred to smaller models ($<$30B parameters) using distillation~\cite{srivastava2025towards}.
	Despite this, most novel techniques still test their effectiveness only on large models (e.g. GPT-4 or GPT-turbo-3.5)~\cite{yao2023tree}~\cite{besta2024graph}~\cite{yuan2025reversal}.

	\subsection{External tools}
	
	Another interesting approach to improving the code quality is to leverage external tools, particularly Static Analysis tools, which can be useful to catch errors in the generated code~\cite{jaoua2025combining}~\cite{blyth2025static}.
	If unit tests are available, the code output can give insights on both syntactic and semantic correctness of the code at the same time.
	For example, syntax errors may include spelling mistakes, undefined functions or variables.
	On the other hand, semantic errors may include incorrect or incomplete logic, which can lead to a code that technically runs without any issues, but returns an incorrect result.
	The code output, whether is an error warning or the result of running unit tests, can be used as feedback to help the model identify issues and fix the code.
	
	Once the code runs without errors and passes all the unit tests, the next step will be to include feedback from Dynamic Analysis tools for time-performance optimization (see Fig.~\ref{fig:agent-workflow}).
	This step is much more complex than verifying code correctness, because the layout can strongly influence runtime and small changes, both related and unrelated to the code, can modify the layout~\cite{mytkowicz2009producing}.
	To address this, we are planning to include tools such as Coz~\cite{coz} that will randomize the layout to ensure a fair comparison between different versions of the code and at the same time to help the model understand which sections of the code have the largest impact.
	We are planning to include remarks from missed compiler optimization for code written in compiled languages (such as C/C++, Rust, etc...) to inform the model of what changes can improve the code.
	We are also considering including performance counters for architecture-dependent optimization, however selecting the best performance counters is not straightforward~\cite{cavazos2007rapidly} and will require a thorough analysis.

	\section{Methodology}
	\label{methodology}

	Motivated by the need to improve the performance of small language models without relying solely on increasing model size, this section evaluates two complementary approaches.
	The first experiment examines repeated sampling under a fixed generation-time budget, while the second examines the use of compiler feedback as additional context.
	
	In this paper we focus on two main research questions:
	\begin{enumerate}
		\item Can multiple runs of a small language model match the performance of a single larger model run? Can this be done with equal or fewer resources?
		\item Can the use of compiler feedback as context improve the performance of small language models?
	\end{enumerate}

	\subsection{Small multi-sampling generation vs large single generation} \label{Jingbo-setup}
	
	To investigate the first research question, we compared repeated sampling of Qwen2.5-Coder-7B with single generation using Qwen2.5-Coder-32B.
	We evaluated the models on CrossCodeEval~\cite{ding2023crosscodeeval}, a multilingual benchmark for repository-level cross-file code completion, in the \texttt{rg1} setting with cross-file context retrieved using UniXCoder similarity.
	We evaluated both models on Python and C\#, using 10 seeds to construct 10 groups of 150 randomly sampled tasks for each programming language.
	For each seed, both models were evaluated on the same group of tasks, enabling paired comparisons.

	\begin{figure}[ht] 
		\includegraphics[width=0.485\textwidth]{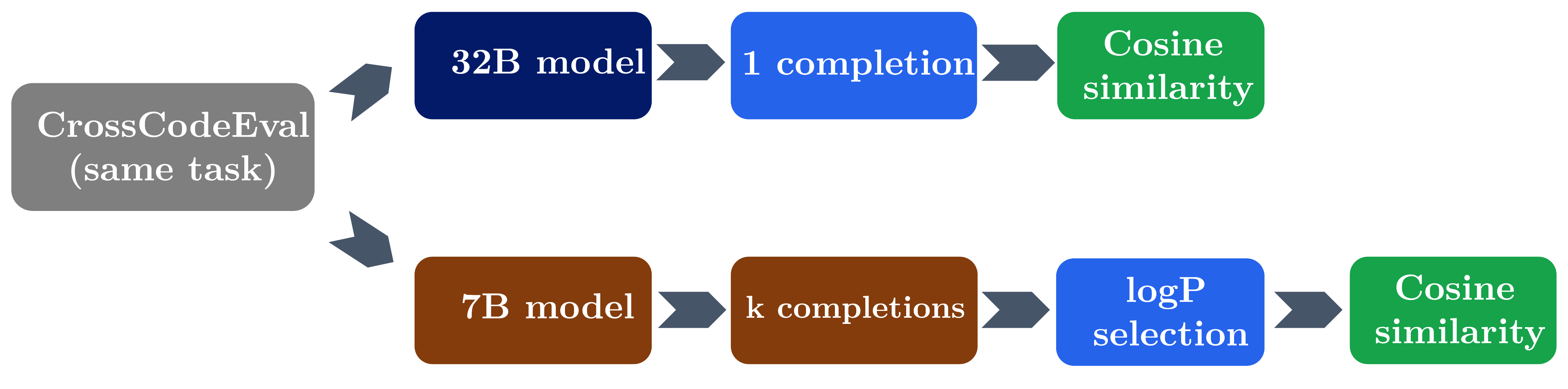}
		\caption{Workflow for the multi-sampling generation vs large single generation experiment. Both systems receive the same task and are compared under the same calibrated generation-only runtime budget. }
		\label{fig:sampling-workflow}
	\end{figure}
	
	For each task, the 32B model generates a single completion, which is compared against the ground truth.
	The 7B model, on the other hand, generates $k$ completions, then we test a naive selection strategy where the model itself selects the completion with the highest log probability ($logP$), which is finally compared against the ground truth (see Fig.~\ref{fig:sampling-workflow}). 
	We used character n-gram cosine similarity to check if the completion is similar to the ground truth.
	Values close to $+1$ indicate high similarity.
	We also evaluated the best-cosine upper bound by comparing all the candidates from the smaller model against the ground truth and selecting the one  with the highest cosine similarity.
	The cosine upper bound represents potential for improvement in the selection strategy, but since it requires ground truth solutions, it can only be applied if unit tests are available.
	
	To evaluate the time budget, we considered generation-only wall-clock time, therefore excluding model loading, environment initialization and file I/O.
	We used one NVIDIA A100-SXM4-40GB GPU for the 7B model and four NVIDIA A100-SXM4-40GB GPUs for the 32B model.
	This was due mainly to memory requirements: unlike the 7B model, the 32B model did not fit in a single GPU.
	Note that this way, our proposed approach is at a disadvantage since the smaller models have fewer computational resources available.
	The inference time of the 32B model defined the time budget, we used a hard threshold for 7B model with multi-sampling generation.
	Consequently, the number of candidate solutions generated by the 7B model was calibrated not to exceed this budget: we used \(k=7\) candidates for Python tasks and \(k=6\) for C\#.

	\subsection{Agentic AI method: Using compiler errors as feedback} \label{agent-setup} 
	\begin{figure*}[ht] 
		\centering
		\includegraphics[width=0.7\textwidth]{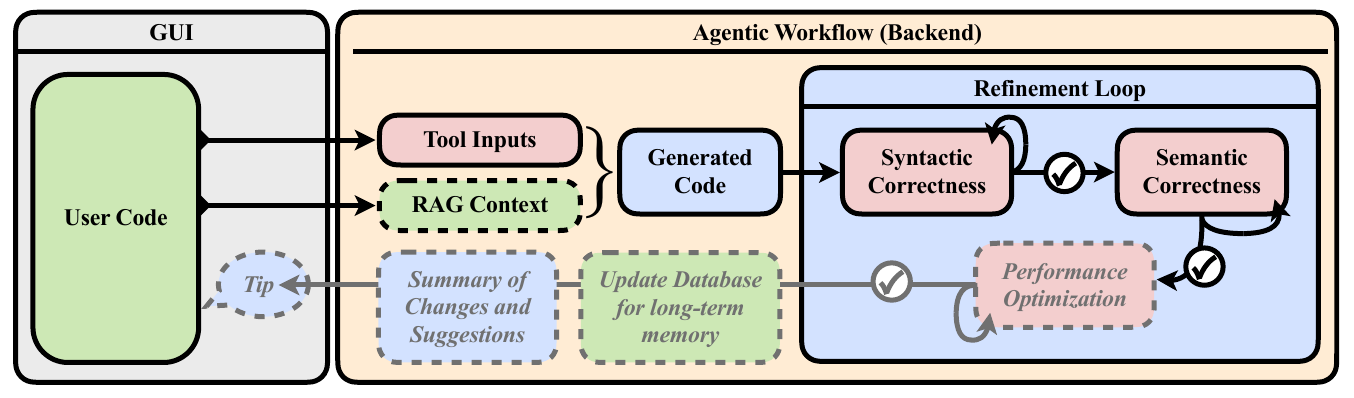}
		\caption{Graphical representation of the agentic workflow. Elements with the gray dashed line are not implemented yet, elements with the black dashed line are implemented but not relevant for the experiments.}
		\label{fig:agent-workflow}
	\end{figure*}

	Another approach to address the fact that smaller models tend to have lower performance than larger models, is to use an agentic workflow.
	We implemented this to give the LMs access to the existing codebase and domain knowledge via RAG, along with feedback from tools integrated in a revision loop (see Fig.~\ref{fig:agent-workflow}), which will be used to test the model's suggestions and reach better performance.
	Among the tools, we included a tool to run the code (either an interpreter or a compiler) to check syntactic correctness and, if unit tests are provided, it will also check semantic correctness.
	Note that the framework is generic; we can also include more advanced static and dynamic analysis tools.
	In this paper, we demonstrate the feasibility of this method by using only a naive implementation, which already increases performance.
	
	For this experiment, we used an AMD EPYC 9255 system with 48 cores and 768 GiB memory, with no GPUs.
	
	\subsubsection{Base models}
	We used 3 models from the Qwen family and 2 models from other families as a control group:
	\begin{enumerate}
		\item Qwen2.5-Coder-3B-Instruct
		\item Qwen3-4B
		\item Qwen3-Coder-Next (Mixture of Experts: 80B-A3B)
		\item CodeLlama-7b-Instruct-hf:
		\item CodeGemma-7b-it
	\end{enumerate}
	
	All models were imported from Hugging Face using the Transformers library.
	
	There are two main roles in the agent: one is the ``coder'', which will write that baseline code when the code is not provided, and one is the ``debugger'', which will try to fix the code if it returns an error (either due to syntax or failed unit tests).
	We explored the case where the same model is assigned to both roles with separate memory and the case where each role corresponds to a different model.
	
	\subsubsection{RAG and tools}
	RAG was implemented using a naive approach where documents are chunked at fixed length with overlap using Langchain~\cite{langchain} and FAISS~\cite{douze2024faiss}, the latter in particular was used for embedding and similarity search.
	A more efficient version would be contextual RAG, which will be implemented in the future.
	The tool for running the code was implemented using ``subprocess.run'', which is also a naive implementation and will be generalized in the future.
	
	\subsubsection{Reasoning and prompting}
	We note that models from the Qwen family are distilled from DeepSeek~\cite{qwen2024qwen2}~\cite{cao2026qwen3}, therefore they have better reasoning abilities and performances than models with a similar size trained from scratch~\cite{srivastava2025towards}.
	We used different prompts for each role, the ``debugger'' role has the explicit instruction to analyze the logic of the code.
	
	We are also planning to implement a long-term memory, similar to Buffer-of-Thoughts~\cite{yang2024buffer}, to store the reasoning behind the most effective optimization changes.
	Once an optimal solution is reached, the model will extract the key points for suggestions and save the code of the optimal solution in a vector database, accessible via RAG.
	This way, the model can reference the code and reasoning if it encounters a similar issue in the future.
	
	\subsubsection{Ideas for the optimization step}
	As mentioned in Section \ref{intro}, we will not focus on optimization in this paper, however we want to outline some possible solutions we will test in future work.
	We are planning to include feedback from Coz~\cite{coz} during the optimization step, as well as the missed remarks from the compiler (only for compiled languages, such as C/C++).
	These remarks are produced by the compiler, similarly to errors and warnings, and they will inform the LM of what changes in the code can result in better performances based on missed optimizations.
	Our main reason for choosing Coz, instead of other dynamic analysis tools, is because Coz removes layout-driven variance and can estimate the optimization potential of different sections in the code using progress points~\cite{curtsinger2015coz}.
	
	\subsubsection{Sustainability metric}
	Since the end-goal of this work is to reduce the environmental impact of large-scale science instruments, the resulting code should be analyzed using a sustainability metric based on energy, carbon and water emissions at job level.
	However, the definition and evaluation of this metric are outside the scope of this paper.
	Initial work in this direction is reported in Chen et al. 2026 \cite{chen2026joblevel}.

	\subsection{Adaptations for benchmarking}
	To test against conventional benchmarks, the agentic workflow needs to be adjusted: the model is meant to give suggestions based on existing code, not write code from scratch.
	To make a meaningful comparison and remain faithful to the intended use case, we first ask the model to generate a baseline code, without relying on any of the tools, and to include assert statements to verify that the function returns the expected result for all the examples given in the prompt which serve the role of unit tests.
	Then, we run the code, collect compiler errors, and ask the model to improve the initial solution, given the prompt, the baseline code and the feedback from the compiler.
	
	We noted that in function-level benchmarks, the generated code would naturally include only the definition of the function, but the function itself is never called, meaning that the compiler would not return feedback on whether the function is working as intended or not.
	To counter this, both the baseline and the final code should test the function on the examples present in the prompt.
	However, it is difficult to extract the examples automatically due to their non-uniform format.
	Therefore, we included a generic prompt before each task with the explicit instruction to write the function and to include the examples as unit tests using assert statements (some models tend to explain the function instead of completing the code).
	
	This means that occasionally the model may hallucinate extra examples with an incorrect canonical solution.
	In this case, the code is considered incorrect even if the generated code would predict the correct solution.
	We also note that the code is considered correct only if all the unit tests and all items in the validation set return the expected result, which means that we are studying the lower-bound performance of the models.
	
	One final point is that conventional benchmarks don't rely on domain knowledge, so retrieving information via RAG is unnecessary to evaluate the performance on benchmarks.
	
	We tested the effectiveness of error feedback on the HumanEval benchmark~\cite{chen2021codex} with 5 samples per task for all models.
	For each model, we evaluate the pass rate with:
	\begin{enumerate}
		\item the \textbf{baseline} mode, which corresponds to the ``coder assistant'' role, where the model receives only the prompts;
		\item the \textbf{agentic} mode, where the ``debugger assistant'' receives the prompt, the code from the coder assistant and compiler errors.
	\end{enumerate}
	
	Moreover, we evaluate the performance in two cases: the code includes only the unit tests, or the code includes both unit tests and validation set. It's worth mentioning that most models were tested by assigning the same model to both roles (coder and debugger), however the last 2 models used Qwen2.5-Coder-3B-Instruct as the coder and Qwen3-4B with thinking enabled as debugger, with either 1 or 3 iteration over the same code.
	
	The reasoning behind this hybrid-role experiment comes from the observation (details in Section \ref{error-analysis}) that the main cause of errors in Qwen2.5-Coder-3B-Instruct was incorrect logic, while for Qwen3-4B the main source of errors was thinking loops or syntax errors.
	However, we observed that this hybrid approach doesn't yield better performance than using the same model for both roles, the performance tends to be aligned with that of the model used for the debugger role, while the main source of errors in hybrid mode shifts back to incorrect logic.
	We also observe no significant difference between 1 and 3 iterations, therefore we conclude that asking the model to fix the same issue multiple times is not an effective strategy.

	\section{Results} \label{results}
	\subsection{Small multi-sampling generation vs large single-generation} \label{Jingbo-results}
	In this section, we evaluate whether repeated sampling with Qwen2.5-Coder-7B can match or outperform a single generation from Qwen2.5-Coder-32B under the same calibrated generation-only time budget.
	We compare both the runtime and the similarity of generated code completions against the ground truth for Python and C\#, across 10 groups with 150 randomly selected tasks each.
	
	\begin{figure}[ht] 
		\includegraphics[width=0.48\textwidth]{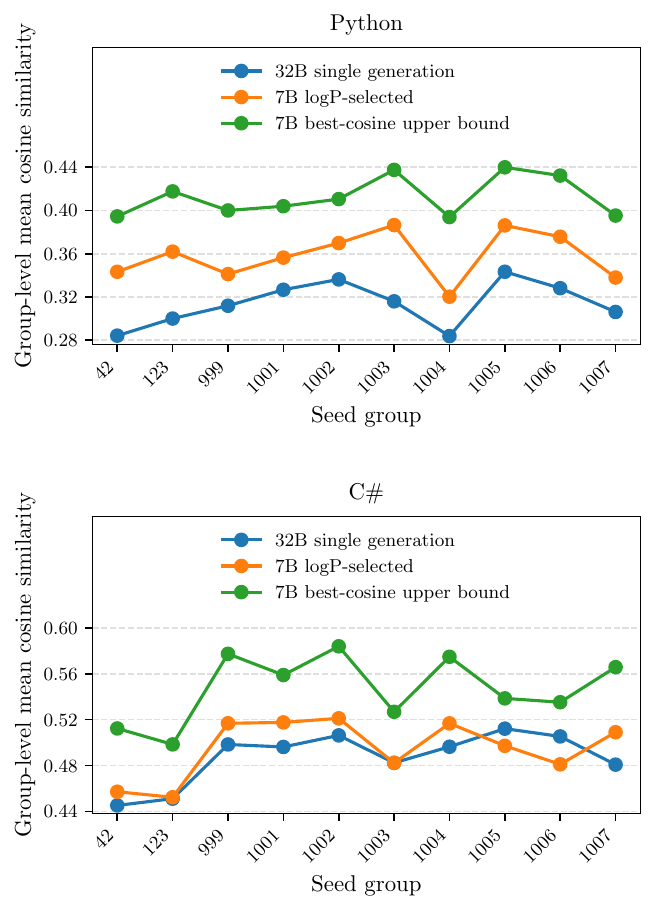}
		\caption{Group-level mean cosine similarity across ten seed groups for Python and C\#. Each seed defines a different randomly sampled task group, so variation across seeds reflects sensitivity to task selection.}
		\label{fig:seeds-performance}
	\end{figure}
	We observe that in Python the 7B model outperforms the 32B model on all groups, indicating low sensitivity to task selection (see Fig.~\ref{fig:seeds-performance}).
	The paired \(t\)-test yielded \(p=0.000006\), the 95\% confidence interval (CI) is \([0.033649, 0.054992]\), indicating a statistically significant improvement.
	
	In C\# the 7B model slightly outperforms the 32B model in 6 tasks, matches the performances in 2, and underperforms in 2,  indicating greater sensitivity to task selection.
	The paired \(t\)-test yielded \(p=0.184941\), the 95\% CI is \([-0.004436, 0.019845]\), indicating the difference was not statistically significant.
	However, in both Python and C\#, the best cosine upper bound shows potential for improvement in the selection strategy.
	
	\begin{figure}[ht] 
		\includegraphics[width=0.48\textwidth]{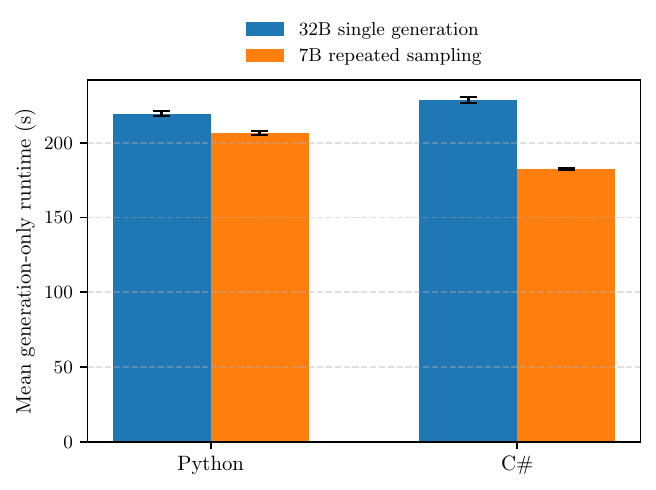}
		\caption{Mean generation-only runtime across ten seed groups for 32B single generation and 7B repeated sampling in Python and C\#. Error bars show standard deviations.}
		\label{fig:runtime}
	\end{figure}
	The 7B model is constrained to have a lower generation-only runtime than the 32B model, and it shows very little variance across the 10 groups (see Fig.~\ref{fig:runtime}). 

	Overall, we show that the 7B model achieved equivalent of higher quality code, at a lower generation-only runtime, while using only one NVIDIA A100-SXM4-40GB GPU compared with 4 GPUs for the 32B baseline.
	Taken together, the lower runtime and lower GPU count strengthen the evidence that repeated sampling can make a smaller model competitive with a larger same-family model.

	\subsection{Agentic AI method: Using compiler errors as feedback} \label{agent-results}
	In this section, we evaluate the effect of compiler feedback on the models' performance, analyzing models from different families with different sizes and architectures.
	
	\begin{figure}[ht] 
		\includegraphics[width=0.485\textwidth]{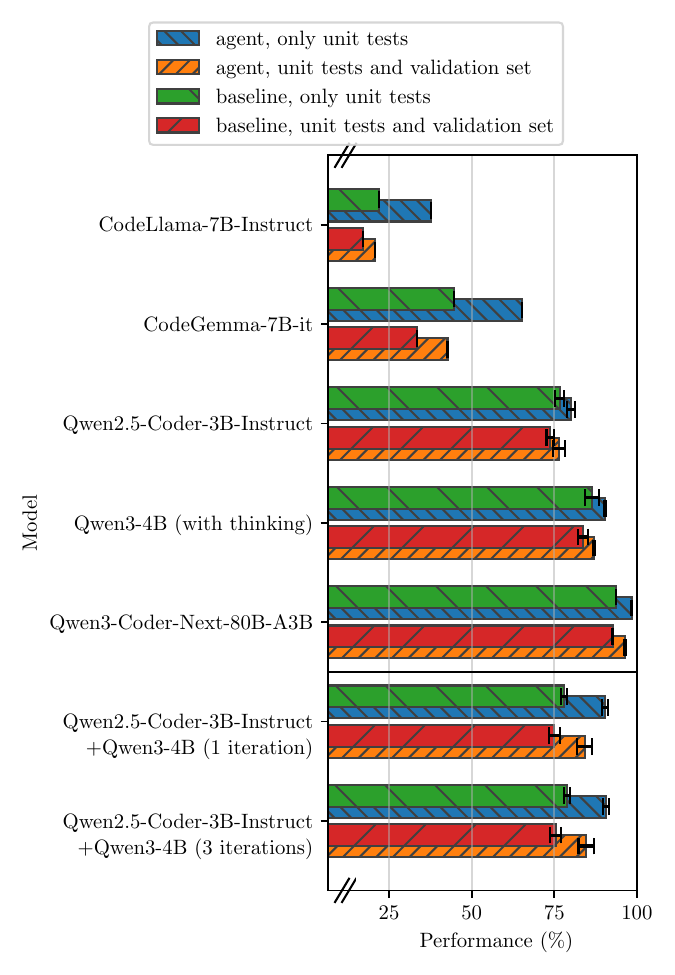}
		\caption{Performance for different models (ordered by performance). The subset of models on the top uses the same model for both the ``coder'' and the ``debugger'' role, while the subset at the bottom are hybrids where we used different models: the first for the coder role and the second for the debugger role. The number of iterations in the second subset represents the max number of times the debugger can iterate on a task.}
		\label{fig:agent-performance}
	\end{figure}
	
	\begin{figure}[ht] 
		\includegraphics[width=0.485\textwidth]{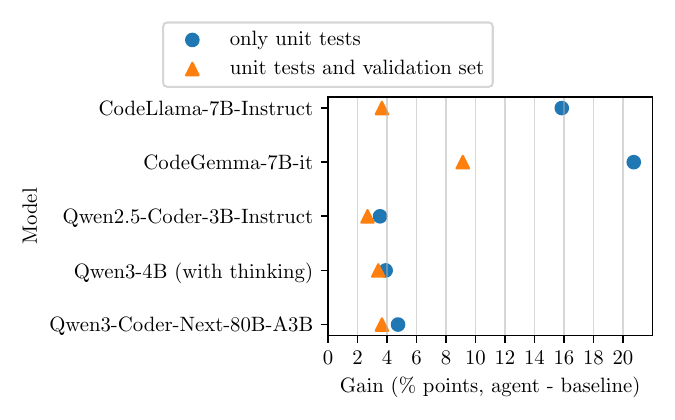}
		\caption{Performance gain for different models (ordered by baseline performance). The plot reports the gain in percentage points of agent minus baseline. The performance gain on the tests appears to be mostly independent from the performance of the base model.}
		\label{fig:gain}
	\end{figure}
	
	We observe that the performance is increased consistently for all models when including compiler errors as feedback (see Fig.~\ref{fig:agent-performance}).
	The performance gain on the validation set is around 3-4 percentage points for most models and seems to be independent from the baseline performance (see Fig.~\ref{fig:gain}), with the notable exception of CodeGemma, whose performance improved by 10 percentage points.
	
	Moreover, we note that the feedback given to the model is limited to the unit tests: feedback from validation set is not included unless the same item is also present in the prompt.
	We observe that in all models the performance on the validation set is increased, suggesting that the unit tests can help the model write a better function, which is more likely to pass the validation set, which may also include edge cases.
	
	We observe that in the Qwen family, the performance gain observed by testing the model only on the unit tests is similar to the one observed when testing both unit tests and validation set, while in Llama and Gemma there is a considerable gap. 
	Qualitative observations suggest that these models may have included some ad-hoc solutions to correctly predict specific unit tests (i.e., they are overfitting), rather than generating a more general solution.

	\subsection{Analysis of errors in the generated code} \label{error-analysis}
	In this section, we report the main reasons why different models generate code that produced incorrect results for at least one of the unit tests present in the prompt, despite being given the compiler errors as feedback.
	We only did this analysis for some of the models (specifically those with fewer errors) because it is a very time-consuming process that requires analyzing each incorrect code completion and its canonical solution to determine the cause of failure. 
	
	We considered the following categories of errors:
	\begin{itemize}
		\item syntax error in the assert statement;
		\item hallucinated extra examples with incorrect solution;
		\item the code was incomplete;
		\item numerical fluctuations of the result;
		\item partially incorrect logic of the solution;
		\item misuse of python functions;
		\item infinite thinking loop, resulting in no code.
	\end{itemize}
	
	\subsubsection{Qwen2.5-Coder-3B-Instruct}
	The main source of errors is partially incorrect logic (50\% of the errors), where the model fails to capture nuances or implicit instructions of the prompt.
	The second most common source of errors is due to hallucinated examples (20\% of the errors).
	
	\subsubsection{Qwen3-4B with thinking enabled}
	Here the main source of errors is incorrect syntax (40\% of the errors), followed by infinite thinking loop (20\% of the errors) and incorrect logic (20\% of the errors).
	
	\subsubsection{Hybrid mode}
	Using the Qwen2.5-Coder-3B-Instruct first as the coder assistant and then Qwen3-4B with thinking as the debugger assistant, the main source of errors returns to be incorrect logic (44\% of the errors), followed by hallucinated examples and misuse of Python (both 19\% of the errors).
	
	\subsubsection{Qwen3-Coder-Next with thinking enabled}
	Here the main source of errors is incorrect logic (70\% of the errors), followed by infinite thinking loops (23\% of the errors).

	\section{Discussion} \label{discussion}
	In this work, we explore two orthogonal solutions to enhance SLMs, both produced positive results, we have not evaluated the combinations yet.
	The two approaches are completely independent, so we expect that when combined we will see the benefits from both.
	
	We had evidence from literature that SLMs could match or outperform larger models on function-level tasks with the same compute budget, but we did not expect to reach the same result while also using fewer computation resources.
	
	We also expected to see some correlation of performance gain with the baseline performance, however the performance gain appears to be mostly independent from model size, class or family for the models we tested.
	Further experimentation is needed to confirm this observation. 
	
	Overall, these results show the potential of SLMs for our application as a valid choice that limits compute budget, and indirectly the carbon footprint, while enhancing the quality of the generated code.

	\section{Conclusions \& Future directions} \label{conclusions}
	SuperRAG combines RAG, reasoning and external tools to enhance Small Language Models enough to make them competitive against much larger and resource-intensive LLMs.
	
	We show that SLMs have the potential of matching larger models when given similar computational time: the 7B model matches or outperforms the 32B model in code completion over repository-level tasks in Python and C\# using one quarter of the resources.
	
	We also demonstrated that using compiler errors as feedback consistently increased the performance for all models by at least 3-4 percentage points, even with a naive implementation.
	
	We will next focus on including feedback to improve code optimization.
	We plan to expand the evaluation to multiple programming languages, we're particularly interested in Python and C/C++.
	We plan to include feedback from Coz to randomize the layout and avoid related bias, as well as missed compilers' remarks for compiled languages such as C/C++.

	\section*{Acknowledgment}
	
	This work was funded by the OTP research project "SuperCode: SUstainability PER AI-driven CO-DEsign", project number 2025/TTW/01878134, which is financed by the Dutch Research Council (NWO) domain Applied and Engineering Sciences (TTW).
	
	\bibliographystyle{IEEEtran}
	\bibliography{bibliography}

\end{document}